\documentclass{cjp}
\usepackage{epsfig}

\usepackage{color}

\usepackage[colorlinks,%
            citecolor=red,linkcolor=blue,hypertex, %
            breaklinks=true]{hyperref}
\begin{document}
\title{On the moment of inertia of PSR J0348+0432}
\author{Xian-Feng Zhao$^{1,2 \ast}$
\\ \emph{$^{1}$ School of Sciences, Southwest Petroleum University, Chengdu, 610500, China\\
$^{2}$ School of Electronic and Electrical Engineering, Chuzhou University, Chuzhou, 239000, China}
}
\maketitle
\begin{abstract}
The moment of inertia of the massive neutron star PSR J0348+0432 is studied in the framework of the relativistic mean field theory by choosing suitable hyperon coupling constants. By this method, we find that the suggested radius of the massive neutron star PSR J0348+0432 is in the range $R=12.957\sim12.246$ km by the observation $M$=1.97$\sim$2.05 M$_\odot$. We also find that the suggested moment of inertia $I$ of the massive neutron star PSR J0348+0432 is in the range $I$=1.9073$\times$10$^{45}\sim$1.5940$\times10^{45}$ g.cm$^{2}$ by the observation $M$=1.97$\sim$2.05 M$_\odot$. Massive pulsars hint that the interaction inside them should be very "strong". Though hyperons considered will reduce the maximum mass, but in principle we may have models predicting maximum masses higher than 2 M${_\odot}$ by choosing suitable parameters, in a degree of freedom of hadron. Our calculations have proved the above and perfectly agree with the results both of Aaron W et al and P¨¦tri J et al.
\end{abstract}


\section{Introduction}
In 2013, Antoniadis et al observed the massive neutron star PSR J0348+0432, which is a $2.01\pm0.04$ M$_{\odot}$ pulsar and spins at 39 ms in a 2.46-hour orbit with a $0.172\pm0.003$ M$_{\odot}$ white dwarf~\cite{Antoniadis 2013}. It has highly rotating speed and may be the largest mass neutron star by now. So, how to theoretically describe its mass and its rotational property, e.g. moment of inertia, would be a very important work for astrophysics.

Overgard T et al examined three models of astrophysical quark matter in 1991. Assuming the validity of Einstein's theory of gravitation, they calculated the total mass, radius, and moment of inertia for quark stars~\cite{Overgard 1991}. An approximation for the moment of inertia of a neutron star, in terms of only its mass and radius, is presented by Bao G et al in 1994, and insight into it is obtained by examining the behavior of the relativistic structural equations~\cite{Bao 1994}.

In 2000, Kalogera V et al calculate new optimal bounds on the masses, radii, and moments of inertia of slowly rotating neutron stars that show kilohertz quasi-periodic oscillations (QPOs)~\cite{Kalogera 2000}. Considering the limits derived for the Crab pulsar, moments of inertia for neutron and strange stars were studied by Bejger M et al in 2002~\cite{Bejger 2002}. Lattimer J M et al and Bejger M et al in 2005~\cite{{Lattimer 2005},{Bejger 2005}} and Morrison I A et al in 2014~\cite{Morrison 2014} estimated the constraint on the equation of state with the moment of inertia measurements of the neutron star PSR J0737-3039A.

Wen D H et al calculated the frame dragging effect on moment of inertia and radius of gyration of neutron star in 2007~\cite{Wen 2007}. In 2008, Aaron W et al studied the nuclear constraints on the momenta of inertia of neutron stars~\cite{Aaron 2008}. In 2010, Yunes N et al's results show that the Chern-Simons correction affects only the gravitomagnetic sector of the metric to leading order, thus introducing modifications to the moment-of-inertia but not to the mass-radius relation~\cite{Yunes 2010}. Sensitivity of the moment of inertia of neutron stars to the equation of state of neutron-rich matter was examined using accurately calibrated relativistic mean-field models by Fattoyev F J et al in 2010~\cite{Fattoyev 2010}.

Constraining the mass and moment of inertia of neutron stars from quasi-periodic oscillations in X-ray binaries was examined by P¨¦tri J in 2011~\cite{P¨¦tri 2011}. In 2015, crust thicknesses, moments of inertia and tidal deformabilities was determined by Steiner A W et al with neutron star observations~\cite{Steiner 2015}.


Studies have shown that the mass of a neutron star is very sensitive to the nucleon coupling constants and the hyperon coupling constants~\cite{Glendenning 1985,Glendenning 1991} and the ratio of hyperon coupling constant to nucleon coupling constant is determined in the range of $\sim$ 1/3 to 1~\cite{Glendenning 1991}.



Hartle et al derived the equations of the moment of inertia of a slowly rotating neutron star from the general relativistic theory in 1967~\cite{Hartle 1967} and they calculated the equilibrium structure of rotating white dwarfs and neutron stars in 1968~\cite{Hartle 1968}. In the last few years, although many works have been done on the moment of inertia of neutron stars~\cite{Link 1999,Li 2001,Ryu 2012}, none of them are on the massive neutron star PSR J0348+0432.

In this paper, the relativistic mean field ( RMF ) theory, which can better describe limited nuclear matter~\cite{Zhou03,Zhou10}, is applied to describe the moment of inertia of the massive neutron star PSR J0348+0432 by choosing the suitable hyperon coupling constants.

\section{The RMF theory and the mass of a neutron star}
The Lagrangian density of hadron matter reads as follows~\cite{Glendenning 1997}

\begin{eqnarray}
\mathcal{L}&=&
\sum_{B}\overline{\Psi}_{B}(i\gamma_{\mu}\partial^{\mu}-{m}_{B}+g_{\sigma B}\sigma-g_{\omega B}\gamma_{\mu}\omega^{\mu}
\nonumber\\
&&-\frac{1}{2}g_{\rho B}\gamma_{\mu}\tau\cdot\rho^{\mu})\Psi_{B}+\frac{1}{2}\left(\partial_{\mu}\sigma\partial^{\mu}\sigma-m_{\sigma}^{2}\sigma^{2}\right)
\nonumber\\
&&-\frac{1}{4}\omega_{\mu \nu}\omega^{\mu \nu}+\frac{1}{2}m_{\omega}^{2}\omega_{\mu}\omega^{\mu}-\frac{1}{4}\rho_{\mu \nu}\cdot\rho^{\mu \nu}+\frac{1}{2}m_{\rho}^{2}\rho_{\mu}\cdot\rho^\mu
\nonumber\\
&&-\frac{1}{3}g_{2}\sigma^{3}-\frac{1}{4}g_{3}\sigma^{4}+\sum_{\lambda=e,\mu}\overline{\Psi}_{\lambda}\left(i\gamma_{\mu}\partial^{\mu}
-m_{\lambda}\right)\Psi_{\lambda}
.\
\end{eqnarray}
Here, $\Psi_{B}$ is the Dirac spinor of the baryon B, {\color{blue}whose mass} is $m_{B}$. $\sigma, \omega$ and $\rho$ are the field operators of the $\sigma, \omega$ and $\rho$ mesons, respectively. $m_{\sigma}$, $m_{\omega}$ and $m_{\rho}$ are the masses of these mesons and $m_{\lambda}$ expresses the leptonic mass. $g_{\sigma_{B}}$, $g_{\omega_{B}}$ and $g_{\rho_{B}}$ are, respectively, the coupling constants of the $\sigma$, $\omega$ and $\rho$ mesons with the baryon B. $\frac{1}{3}g_{2}\sigma^{3}+\frac{1}{4}g_{3}\sigma^{4}$ expresses the self-interactive energy, in which $g_{2}$ and $g_{3}$ are the self-interaction parameters of $\sigma$-meson. And the last term expresses the Lagrangian of both electron and muon.

The energy density $\varepsilon$, pressure $p$, mass $M$, radius $R$ and the moment of inertia $I$ of a neutron star can be seen in Ref.~\cite{Glendenning 1997}.

\section{Parameters}
The GL85 nucleon coupling constant can better describe the properties of neutron stars and so we choose it in this work: the saturation density $\rho_{0}$=0.145 fm$^{-3}$, binding energy B/A=15.95 MeV, a compression modulus $K=285$ MeV, charge symmetry coefficient $a_{sym}$=36.8 MeV and the effective mass $m^{*}/m$=0.77~\cite{Glendenning 1985}.

For the hyperon coupling constant, we define the ratios: $x_{\sigma h}=\frac{g_{\sigma h}}{g_{\sigma}}=x_{\sigma}
$, $x_{\omega h}=\frac{g_{\omega h}}{g_{\omega}}=x_{\omega}$, $x_{\rho h}=\frac{g_{\rho h}}{g_{\rho}}$. Here, $h$ denoting hyperons $\Lambda, \Sigma$ and $\Xi$.

We choose $x_{\rho \Lambda}=0, x_{\rho \Sigma}=2, x_{\rho \Xi}=1$ according to SU(6) symmetry at first~\cite{Schaffner 1996}.

For the parameters $x_{\sigma h}$, we choose $x_{\sigma h}$=0.4, 0.5, 0.6, 0.7, 0.8, 0.9 at first(h denotes hyperons $\Lambda$, $\Sigma$ and $\Xi$, respctively)~\cite{Glendenning 1991}. For each $x_{\sigma h}$, the $x_{\omega h}$ is chosen to fit to the hyperon well depth~\cite{Glendenning 1997}

\begin{eqnarray}
U_{h}^{(N)}=m_{B}\left(\frac{m_{n}^{*}}{m_{n}}-1\right)x_{\sigma h}+\left(\frac{g_{\omega N}}{m_{\omega}}\right)^{2}\rho_{0}x_{\omega h}
.\
\end{eqnarray}
The experimental data of the well depth are $U_{\Lambda}^{(N)}=-30$ MeV~\cite{Batty 1997}, $ U_{\Sigma}^{(N)}=10\sim40$ MeV~\cite{Kohno 2004,Kohno 2006,Harada 2005,Harada 2006,Friedman 2007} and $U_{\Xi}^{(N)}=-28$ MeV~\cite{Schaffner-Bielich 2000}. Therefore, in this work we choose $U_{\Lambda}^{(N)}=-30$ MeV, $ U_{\Sigma}^{(N)}=+40$ MeV and $U_{\Xi}^{(N)}=-28$ MeV.

For $x_{\sigma \Lambda}$=0.9, we get $x_{\omega \Lambda}$=1.0729, which is greater than 1 and is deleted. So we only choose $x_{\sigma \Lambda}$=0.4, 0.5, 0.6, 0.7, 0.8 and correspondingly we obtain $x_{\omega \Lambda}$=0.3679, 0.5090, 0.6500, 0.7909, 0.9319, respectively(see Table.~\ref{tab1}).

As $x_{\sigma \Sigma}$=0.6, 0.7, 0.8, 0.9, $x_{\omega \Sigma}$ all will be greater than 1 (e.g. $x_{\sigma \Sigma}$=0.6, $x_{\omega \Sigma}$=1.1069). Thus, we can only choose $x_{\sigma \Sigma}$=0.4, 0.5, correspondingly we obtain $x_{\omega \Sigma}$=0.8250, 0.9660. For the positive $ U_{\Sigma}^{(N)}$ restricting the production of the hyperon $\Sigma$~\cite{Zhao11}, so in this work we only choose $x_{\sigma \Sigma}=0.4$, $x_{\omega \Sigma}$=0.8250 while $x_{\sigma \Sigma}=0.5$, $x_{\omega \Sigma}$=0.9660 can be deleted (see Table~\ref{tab1}).

For $x_{\sigma \Xi}$, {\color{blue}we first choose $x_{\sigma \Xi}$=0.4, 0.5, 0.6, 0.7, 0.8, 0.9. Through fitting to the hyperon well depth, we obtain $x_{\omega \Xi}$=0.3811, 0.5221, 0.6630, 0.8040, 0.9450, 1.0860, respectively.} Obviously, $x_{\sigma \Xi}$=0.9 and $x_{\omega \Xi}$=1.0860 should be deleted. Thus, the parameters that can be chosen are listed in Table~\ref{tab1}.

\begin{table}[!htbp]
\centering
\caption{The hyperon coupling constants fitting to the experimental data of the well depth, which are $U_{\Lambda}^{N}=-30$ MeV, $U_{\Sigma}^{N}=+40$ MeV and $U_{\Xi}^{N}=-28$ MeV, respectively.}
\label{tab1}
\begin{tabular}[t]{llllll}
\hline\noalign{\smallskip}
$x_{\sigma \Lambda}$ &$x_{\omega \Lambda}$&$x_{\sigma \Sigma}$&$x_{\omega \Sigma}$&$x_{\sigma \Xi}$     &$x_{\omega \Xi}$\\
\hline
0.4               &0.3679              &0.4             &0.8250            &0.4               &0.3811   \\
0.5               &0.5090              &\underline{0.5} &\underline{0.9660}&0.5               &0.5221    \\
0.6               &0.6500              &                   &                  &0.6               &0.6630     \\
0.7               &0.7909              &                   &                  &0.7               &0.8040    \\
0.8               &0.9319              &                   &                  &0.8               &0.9450    \\
\noalign{\smallskip}\hline\noalign{\smallskip}
\end{tabular}
\vspace*{0.6cm}  
\end{table}

From the parameters chosen above, we can compose of 25 sets of suitable parameters(named as NO.1-25){\color{blue}(see Table~\ref{tab2})}.

\begin{table}[!htbp]
\centering
\caption{{\color{blue}The 25 sets of hyperon coupling constants used in this work.}}
\label{tab2}
\begin{tabular}[t]{lllllll}
\hline\noalign{\smallskip}
NO.&$x_{\sigma \Lambda}$ &$x_{\omega \Lambda}$&$x_{\sigma \Sigma}$&$x_{\omega \Sigma}$&$x_{\sigma \Xi}$     &$x_{\omega \Xi}$\\
\hline
01&0.4               &0.3679              &0.4             &0.8250            &0.4               &0.3811   \\
02&0.4               &0.3679              &0.4             &0.8250            &0.5               &0.5221    \\
03&0.4               &0.3679              &0.4             &0.8250            &0.6               &0.6630     \\
04&0.4               &0.3679              &0.4             &0.8250            &0.7               &0.8040    \\
05&0.4               &0.3679              &0.4             &0.8250            &0.8               &0.9450    \\
06&0.5               &0.5090              &0.4             &0.8250            &0.4               &0.3811   \\
07&0.5               &0.5090              &0.4             &0.8250            &0.5               &0.5221    \\
08&0.5               &0.5090              &0.4             &0.8250            &0.6               &0.6630     \\
09&0.5               &0.5090              &0.4             &0.8250            &0.7               &0.8040    \\
10&0.5               &0.5090              &0.4             &0.8250            &0.8               &0.9450    \\
11&0.6               &0.6500              &0.4             &0.8250            &0.4               &0.3811   \\
12&0.6               &0.6500              &0.4             &0.8250            &0.5               &0.5221    \\
13&0.6               &0.6500              &0.4             &0.8250            &0.6               &0.6630     \\
14&0.6               &0.6500              &0.4             &0.8250            &0.7               &0.8040    \\
15&0.6               &0.6500              &0.4             &0.8250            &0.8               &0.9450    \\
16&0.7               &0.7909              &0.4             &0.8250            &0.4               &0.3811   \\
17&0.7               &0.7909              &0.4             &0.8250            &0.5               &0.5221    \\
18&0.7               &0.7909              &0.4             &0.8250            &0.6               &0.6630     \\
19&0.7               &0.7909              &0.4             &0.8250            &0.7               &0.8040    \\
20&0.7               &0.7909              &0.4             &0.8250            &0.8               &0.9450    \\
21&0.8               &0.9319              &0.4             &0.8250            &0.4               &0.3811   \\
22&0.8               &0.9319              &0.4             &0.8250            &0.5               &0.5221    \\
23&0.8               &0.9319              &0.4             &0.8250            &0.6               &0.6630     \\
24&0.8               &0.9319              &0.4             &0.8250            &0.7               &0.8040    \\
25&0.8               &0.9319              &0.4             &0.8250            &0.8               &0.9450    \\
\noalign{\smallskip}\hline\noalign{\smallskip}
\end{tabular}
\vspace*{0.6cm}  
\end{table}

\begin{figure}[!htp]
\centering{}\includegraphics[width=3in]{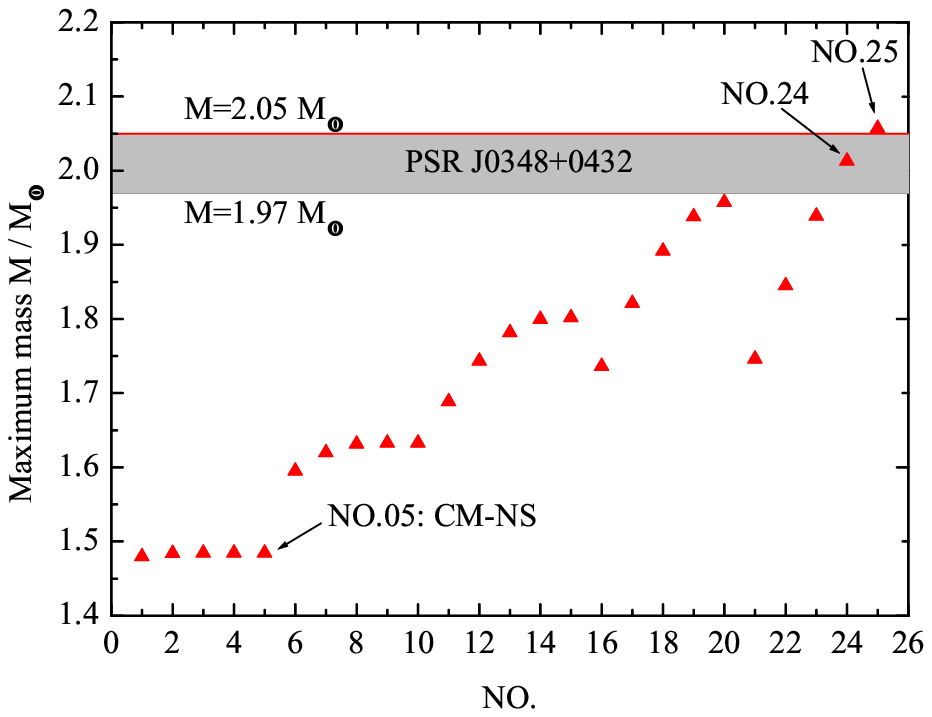}\caption{The mass $M$ of the neutron stars as a function of the central energy density $\varepsilon_{c}$.}
\label{fig1}
\end{figure}

For every set of parameters we calculate the mass of the neutron star. Only parameters NO.24 ( $x_{\sigma \Lambda}=0.8, x_{\omega \Lambda}$=0.9319; $x_{\sigma \Sigma}=0.4, x_{\omega \Sigma}=0.825$; $x_{\sigma \Xi}=0.7, x_{\omega \Xi}=0.804$. $M_{max}$=2.0132 M$_{\odot}$ ) and NO.25 ( $x_{\sigma \Lambda}=0.8, x_{\omega \Lambda}$=0.9319; $x_{\sigma \Sigma}=0.4, x_{\omega \Sigma}=0.825$; $x_{\sigma \Xi}=0.8, x_{\omega \Xi}=0.945$. $M_{max}$=2.0572 M$_{\odot}$ ) can give the mass greater than that of the massive neutron star PSR J0348+0432( see Fig.~\ref{fig1}). Next, we use parameters NO.25 to describe the properties of the neutron star PSR J0348+0432.

{\color{blue}
Considering the SU(6) symmetry~\cite{Schaffner 1996} and choosing $U_{\Lambda}^{(N)}=-30$ MeV, $ U_{\Sigma}^{(N)}=+40$ MeV and $U_{\Xi}^{(N)}=-28$ MeV, the hyperon coupling constants obtained are $x_{\sigma \Lambda}=0.6118$, $x_{\omega \Lambda}$=0.6667; $x_{\sigma \Sigma}=0.2877$, $x_{\omega \Sigma}$=0.6667; $x_{\sigma \Xi}$=0.6026, $x_{\omega \Xi}=0.3333$. In this case, the maximum mass of the neutron star calculated are only $M_{max}$=1.3374 M$_{\odot}$, which is far less than that of the massive neutron star PSR J0348+0432. In order to obtain the larger neutron star mass, we have to broke the SU(6) symmetry and choose parameters NO.25 to describe the neutron star PSR J0348+0432.
}

\section{The radius of the massive neutron star PSR J0348+0432}
\begin{figure}[!htp]
\centering{}\includegraphics[width=3in]{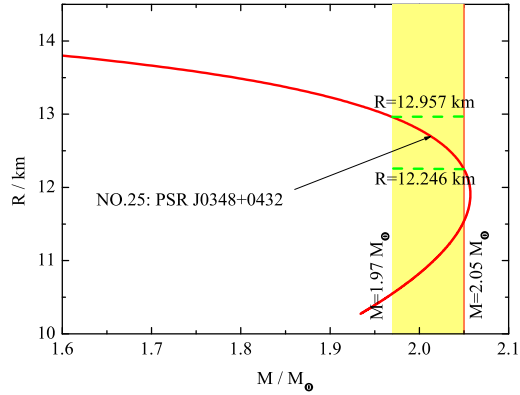}\caption{The radius $R$ of the neutron star as a function of the mass.}
\label{fig2}
\end{figure}

Fig.~\ref{fig2} gives the radius of the neutron star as a function of the mass. We see the radius of the massive neutron star PSR J0348+0432 is $R=12.957$ km as the mass $M$=1.97 M$_\odot$ and is $R=12.246$ km as the mass $M$=2.05 M$_\odot$. That is to say, the suggested radius of the massive neutron star PSR J0348+0432 is in the range $R=12.957\sim12.246$ km by the observation $M$=1.97$\sim$2.05 M$_\odot$. {\color{blue}This result also can be seen} in Table~\ref{tab3}.

We also see from Fig.~\ref{fig2} that the central energy density of the massive neutron star PSR J0348+0432 is $\varepsilon_{c}=1.2601\times10^{15}$ g.cm$^{-3}$ corresponding to the mass $M$=1.97 M$_\odot$ and is $\varepsilon_{c}=1.7774\times10^{15}$ g.cm$^{-3}$ to the mass $M$=2.05 M$_\odot$, namely, the suggested central energy density of the massive neutron star PSR J0348+0432 is in the range $\varepsilon_{c}=1.2601\times10^{15}$ $\sim$ $1.7774\times10^{15}$ g.cm$^{-3}$ by the observation $M$=1.97$\sim$2.05 M$_\odot$(see Table~\ref{tab3}).

\begin{table}[!htbp]
\centering
\caption{The mass $M$, radius $R$ and the moment of inertia $I$ of the massive neutron star PSR J0348+0432.}
\label{tab3}
\begin{tabular}[t]{llll}
\hline\noalign{\smallskip}
     $M$          &$\varepsilon_{c}$                &$R$      &$I$\\
  M$_{\odot}$            &$\times10^{15}$g/cm$^{3}$&km     &$\times10^{45}$g.cm$^{2}$\\
\hline
1.97&1.2601                          &12.957 &1.9073\\
2.05&1.7774                        &12.246 &1.5940\\
\noalign{\smallskip}\hline\noalign{\smallskip}
\end{tabular}
\vspace*{0.6cm}  
\end{table}

\section{The moment of inertia of the massive neutron star PSR J0348+0432}
\begin{figure}[!htp]
\centering{}\includegraphics[width=3in]{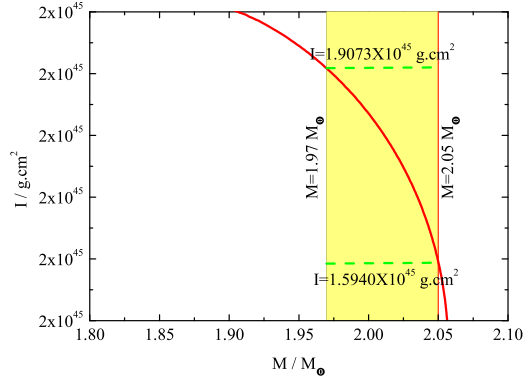}\caption{The moment of inertia $I$ of the neutron star as a function of the mass $M$.}
\label{fig3}
\end{figure}

The moment of inertia $I$ of the neutron star as a function of the mass is given in Fig.~\ref{fig3}, from which we can see that the moment of inertia $I$ decreases with the increase of the mass $M$.

We also see the moment of inertia $I$ of the massive neutron star PSR J0348+0432 is $I$=1.9073$\times10^{45}$ g.cm$^{2}$ corresponding to the mass $M=1.97$ M$_\odot$ and is $I$=1.5940$\times10^{45}$ g.cm$^{2}$ to the mass $M=2.05$ M$_\odot$. The suggested moment of inertia $I$ of the massive neutron star PSR J0348+0432 is in the range $I$=1.9073$\times$10$^{45}\sim$1.5940$\times10^{45}$ g.cm$^{2}$ by the observation $M$=1.97$\sim$2.05 M$_\odot$.

Obviously, other sets of parameters (for example, NO.24) also can be got considering the constraints of the hyperon well depth in the nucleon matter. The coupling constants (NO.25) used in this work is only one of them.

\section{Summary}
In this paper, the moment of inertia of the massive neutron star PSR J0348+0432 is studied in the framework of the RMF theory by choosing suitable hyperon coupling constants. By fitting to the experimental data of the hyperon well depth one model of the massive neutron star PSR J0348+0432 is found. With it the moment of inertia of the massive neutron star PSR J0348+0432 is studied.

We find the suggested radius of the massive neutron star PSR J0348+0432 is in the range $R=12.957\sim12.246$ km and the suggested central energy density is in the range $\varepsilon_{c}=1.2601\times10^{15}$ $\sim$ $1.7774\times10^{15}$ g.cm$^{-3}$ by the observation $M$=1.97$\sim$2.05 M$_\odot$. We also find that the suggested moment of inertia $I$ of the massive neutron star PSR J0348+0432 is in the range $I$=1.9073$\times$10$^{45}\sim$1.5940$\times10^{45}$ g.cm$^{2}$ by the observation $M$=1.97$\sim$2.05 M$_\odot$.

In fact, the variation of the saturation density $\rho_{0}$, binding energy B/A, compression modulus $K$, charge symmetry coefficient $a_{sym}$ and the effective mass $m^{*}/m$ will cause the fluctuation of the maximum mass in the range $1.97\sim3.4$ M$_{\odot}$ and cause the corresponding radius in the range 10.3$\sim15.2$ km with no hypeorns included~\cite{Zhao 2012}. On the other hand, the nucleon coupling constants $g_{\sigma}$, $g_{\omega}$ and $g_{\rho}$ are connected with the properties of the nuclear matter. So, the variation of the nuclear matter would cause the uncertainty of the $g_{\sigma}$, $g_{\omega}$ and $g_{\rho}$. These would further cause the uncertainty of the mass, radius and the moment of inertia. In this work, all these factors are not considered.

Massive pulsars hint that the interaction inside them should be very "strong". Although Hyperons considered will reduce the maximum mass~\cite{Glendenning 1985}, in principle one may have models predicting maximum masses higher than 2 M${_\odot}$ by choosing "suitable" parameters, in a degree of freedom of either hadron or quark.

The results of Aaron W et al show the moment of inertia is in the range of $I$=1.3$\sim$1.63$\times10^{45}$ g.cm$^{2}$ for the neutron star mass 1.338 M$_{\odot}$~\cite{Aaron 2008} and those of P¨¦tri J et al is in the range of $I$=1$\sim$3$\times10^{45}$ g.cm$^{2}$ corresponding to the neutron star mass 2$\sim$2.2 M$_{\odot}$~\cite{P¨¦tri 2011}. Our calculations perfectly agree with them.

\textbf{Acknowledgements}\\
{\color{blue}We are thankful to Shan-Gui Zhou for fruitful discussions during the visit to Institute of Theoretical Physics of Chinese Academy of Sciences and anonymous referee for some valuable suggestions}. This work was supported by the Natural Science Foundation of China ( Grant No. 11447003 ) and the Scientific Research Foundation of the Higher Education Institutions of Anhui Province, China ( Grant No. KJ2014A182 ).

\end{document}